\documentclass[aps,prd,preprint,12pt,superscriptaddress,nofootinbib,a4paper]{revtex4-1}

\usepackage[utf8]{inputenc}
\usepackage{amsmath,amssymb}
\usepackage[separate-uncertainty=true]{siunitx}
\usepackage{graphicx}
\usepackage[usenames,dvipsnames]{xcolor}
\usepackage[caption=false]{subfig}
\usepackage{hyperref}
\usepackage[capitalize]{cleveref}
\usepackage{booktabs}
\usepackage{tabularx}
\usepackage{xspace}
\usepackage{color}
\usepackage[normalem]{ulem}
\usepackage{slashed}
\usepackage{bbold}
\usepackage{wasysym}
\usepackage{graphicx}
\usepackage{mathrsfs} 

\allowdisplaybreaks

\graphicspath{{graphics/}}

\newcommand{\eqdot}{\,.}
\newcommand{\eqcomma}{\,,}
\newcommand{\eqn}{equation}
\newcommand{\lb}{\left(}
\newcommand{\rb}{\right)}

\newcommand{\Ztwo}{\ensuremath{{\mathbb{Z}_2}}\xspace}
\newcommand{\HiggsBounds}{\texttt{HiggsBounds}\xspace}
\newcommand{\HiggsSignals}{\texttt{HiggsSignals}\xspace}
\newcommand{\HSv}[1]{\texttt{HiggsSignals-#1}}
\newcommand{\HBv}[1]{\texttt{HiggsBounds-#1}}

\newcommand{\GeV}{{\ensuremath\rm GeV}}
\newcommand{\TeV}{{\ensuremath\rm TeV}}

\newcommand{\pb}{{\ensuremath\rm pb}}
\newcommand{\fb}{{\ensuremath\rm fb}}

\DeclareSIUnit{\pb}{pb}
\DeclareSIUnit{\fb}{fb}
\AtBeginDocument{
\heavyrulewidth=.08em
\lightrulewidth=.05em
\cmidrulewidth=.03em
\belowrulesep=.65ex
\belowbottomsep=0pt
\aboverulesep=.4ex
\abovetopsep=0pt
\cmidrulesep=\doublerulesep
\cmidrulekern=.5em
\defaultaddspace=.5em

\newcolumntype{C}{>{\centering\arraybackslash}X}
\newcolumntype{b}{C}
\newcolumntype{s}{>{\hsize=.6\hsize}C}
\newcolumntype{R}{>{\raggedleft\arraybackslash}X}
}

\begin{document}
\bibliographystyle{hunsrt}
\date{\today}
\rightline{RBI-ThPhys-2022-21}
\title{{\Large TRSM Benchmark Planes - Snowmass White Paper}}

\author{Tania Robens}
\email{trobens@irb.hr}
\affiliation{Ruder Boskovic Institute, Bijenicka cesta 54, 10000 Zagreb, Croatia}

\renewcommand{\abstractname}{\texorpdfstring{\vspace{0.5cm}}{} Abstract}

\begin{abstract}
    \vspace{0.5cm}
In this whitepaper, I briefly review the Benchmark Planes in the Two-Real-Singlet Model (TRSM), a model that enhances the Standard Model (SM) scalar sector by two real singlets that obey a $\mathbb{Z}_2\,\otimes\,\mathbb{Z}_2'$ symmetry. In this model, all fields acquire a vacuum expectation value, such that the model contains in total 3 CP-even neutral scalars that can interact with each other. All interactions with SM-like particles are inherited from the SM-like doublet via mixing. I remind the readers of the previously proposed benchmark planes, and briefly discuss possible production at future Higgs factories.
  
\end{abstract}

\maketitle


\section{Introduction and Model}
The model discussed here has been proposed in \cite{Robens:2019kga}, and I refer the reader to that reference for a detailed discussion of model setup and constraints. I just briefly repeat the generic features for completeness.

The potential in the scalar sector is given by
\begin{equation}
    \begin{aligned}
        V\lb \Phi,\,S,\,X\rb & = \mu_{\Phi}^2 \Phi^\dagger \Phi + \lambda_{\Phi} {(\Phi^\dagger\Phi)}^2
        + \mu_{S}^2 S^2 + \lambda_S S^4
        + \mu_{X}^2 X^2 + \lambda_X X^2                                              \\
          & \quad+ \lambda_{\Phi S} \Phi^\dagger \Phi S^2
        + \lambda_{\Phi X} \Phi^\dagger \Phi X^2
        + \lambda_{SX} S^2 X^2\eqdot
    \end{aligned}\label{eq:TRSMpot}
\end{equation}
Here, $\Phi$ denotes the SM-like doublet, while $X,\,S$ are two additional real scalar fields. The model obeys an additional $\Ztwo\,\otimes\,\Ztwo'$ symmetry
$        \Ztwo^S: \, S\to -S\eqcomma
        \Ztwo^X: \, X\to -X,$ while all other fields transform evenly under the respective $\Ztwo$ symmetry. All three scalars acquire a vacuum expectation value (vev) and therefore mix. This leads to three physical states with all possible scalar-scalar interactions.

Among the important constraints are e.g. the Higgs signal strength measurements by the LHC experiments, perturbative unitarity as well as the requirement for the potential to be bounded from below, and current collider searches. Results have been obtained using the \texttt{ScannerS}~\cite{Coimbra:2013qq,Ferreira:2014dya,Costa:2015llh,Muhlleitner:2016mzt,Muhlleitner:2020wwk} framework. Experimental results from past and current collider experiments have been implemented using the publicly available tools \HiggsBounds \cite{Bechtle:2008jh,Bechtle:2011sb,Bechtle:2013gu,Bechtle:2013wla, Bechtle:2015pma,Bechtle:2020pkv} and \HiggsSignals \cite{Stal:2013hwa, Bechtle:2013xfa,Bechtle:2014ewa,Bechtle:2020uwn}.

In the following, we will use the convention that
\begin{\eqn}\label{eq:hier}
M_1\,\leq\,M_2\,\leq\,M_3
\end{\eqn}
and denote the corresponding physical mass eigenstates by $h_i$.
Gauge and mass eigenstates are related via a mixing matrix. Interactions with SM particles are then inherited from the scalar excitation of the doublet via rescaling factors $\kappa_i$, such that $g_i^{h_i A B}\,=\,\kappa_i\,g_i^{h_i A B,\text{SM}}$ for any $h_i A B$ coupling, where $A,\,B$ denote SM particles. Orthogonality of the mixing matrix implies $\sum_i \kappa_i^2\,=\,1$. Furthermore, signal strength measurements require $|\kappa_{125}|\gtrsim\,0.96$ \cite{Robens:2019kga} for the SM-like scalar $h_{125}$, which can be $h_1,\,h_2$ or $h_3$ depending on the specific parameter choice.

For a certain production process (e.g.~gluon gluon fusion) the cross
section, $\sigma$, for $h_a$ with mass $M_a$ can be obtained from the
corresponding SM Higgs production cross section, $\sigma_\text{SM}$, by
simply rescaling
\begin{equation}
    \sigma(M_a) = \kappa_a^2 \cdot \sigma_\text{SM} ( M_a)\eqdot\label{eq:cxnscaling}
\end{equation}
Since $\kappa_a$ rescales all Higgs couplings to SM particles,
\cref{eq:cxnscaling} is exact up to genuine electroweak corrections involving Higgs
self-interactions, and in particular holds to all orders in QC\@D.

The scaling factor $\kappa_a$ plays the same role in universally rescaling the partial widths of
$h_a$ decays into SM particles, leading to 
\begin{equation}
    \Gamma(h_a\to\text{SM}; M_a) = \kappa_a^2 \cdot \Gamma_\text{tot}(h_\text{SM}; M_a),\label{eq:widthscaling}
\end{equation}
where $\Gamma( h_a \to \text{SM}; M_a)$ denotes the sum of all partial widths of $h_a$ into SM particle final states. In addition, the branching ratios (BRs) of $h_a$ decays to other scalar bosons, $h_a \to h_b h_c$, are given by:
\begin{equation}
    \text{BR}(h_a\to h_b h_c) = \frac{\Gamma_{a\to bc}}{\kappa_a^2~\Gamma_\text{tot}(h_\text{SM}) + \sum_{xy} \Gamma_{a\to xy}}\eqdot
\end{equation}
where the denominator now denotes the total width of the scalar $h_a$. In the absence of BSM decay modes --- which
is always the case for the lightest Higgs bosons $h_1$ --- $h_a$ has BRs
identical to a SM-like Higgs boson of the same mass.
\section{Benchmark planes}
In \cite{Robens:2019kga}, several benchmark planes were proposed which were meant to capture mainly features that by the time of that publication were not yet adressed by searches at the LHC:
\begin{itemize}
\item{}asymmetric production and decay, in the form of
\begin{\eqn*}
p,\,p\,\rightarrow\,h_3\,\rightarrow\,h_1\,h_2,
\end{\eqn*}
where, depending on the kinematics, $h_2\,\rightarrow\,h_1\,h_1$ decays are also possible;
\item{}symmetric decays in the form of
\begin{\eqn*}
p\,p\,\rightarrow\,h_i\,\rightarrow\,h_j\,h_j,
\end{\eqn*}
where none of the scalars corresponds to the 125 \GeV~ resonance. Note that this in principle allows for further decays $h_j\,\rightarrow\,h_k\,h_k$, again depending on the specific benchmark plane kinematics.
\end{itemize}
We list the definition of these benchmark planes in tables \ref{tab:benchmarkoverview} and \ref{tab:BPparams}, respectively.

\begin{table} 
    \centering
    \begin{tabularx}{\textwidth}{sssb}
        \toprule
        benchmark scenario & $h_{125}$ candidate & target signature      & possible successive decays                          \\
        \midrule
        \textbf{BP1}       & $h_3$               & $h_{125} \to h_1 h_2$ & $h_2 \to h_1 h_1$ if $M_2 > 2 M_1$                  \\
        \textbf{BP2}       & $h_2$               & $h_3 \to h_1 h_{125}$ & -                                                   \\
        \textbf{BP3}       & $h_1$               & $h_3 \to h_{125} h_2$ & $h_2 \to h_{125}h_{125}$ if $M_2 > \SI{250}{\GeV}$  \\
        \textbf{BP4}       & $h_3$               & $h_2 \to h_1 h_1$     & -                                                   \\
        \textbf{BP5}       & $h_2$               & $h_3 \to h_1 h_1$     & -                                                   \\
        \textbf{BP6}       & $h_1$               & $h_3 \to h_2 h_2$     & $h_2 \to h_{125}h_{125}$ if  $M_2 > \SI{250}{\GeV}$ \\
        \bottomrule
    \end{tabularx}
    \caption{Overview of the benchmark scenarios: The second column denotes the
    Higgs mass eigenstate that we identify with the observed Higgs boson,
    $h_{125}$, the third column names the targeted decay mode of the resonantly
    produced Higgs state, and the fourth column lists possible relevant
    successive decays of the resulting Higgs states.}\label{tab:benchmarkoverview}
\end{table}
\begin{table} 
    \centering
    \begin{tabularx}{\textwidth}{CRRRRRR}
        \toprule
        Parameter           & \multicolumn{6}{c }{Benchmark scenario}                                                                             \\
                            & \textbf{BP1}                            & \textbf{BP2} & \textbf{BP3} & \textbf{BP4} & \textbf{BP5} & \textbf{BP6}  \\
        \midrule
        $M_1~[\SI{}{\GeV}]$ & $[1, 62]$                               & $[1,124]$    & $125.09$     & $[1, 62]$    & $[1, 124]$   & $125.09$      \\
        $M_2~[\SI{}{\GeV}]$ & $[1, 124]$                              & $125.09$     & $[126, 500]$ & $[1,124]$    & $125.09$     & $[126, 500]$  \\
        $M_3~[\SI{}{\GeV}]$ & $125.09$                                & $[126,500]$  & $[255, 650]$ & $125.09$     & $[126, 500]$ & $[255, 1000]$ \\
        $\theta_{hs}$       & $1.435$                                 & $1.352$      & $-0.129$     & $-1.284$     & $-1.498$     & $0.207$       \\
        $\theta_{hx}$       & $-0.908$                                & $1.175$      & $0.226$      & $1.309$      & $0.251$      & $0.146$       \\
        $\theta_{sx}$       & $ -1.456$                               & $-0.407$     & $-0.899$     & $-1.519$     & $0.271$      & $0.782$       \\
        $v_s~[\SI{}{\GeV}]$ & $630$                                   & $120$        & $140$        & $990$        & $50$         & $220$         \\
        $v_x~[\SI{}{\GeV}]$ & $700$                                   & $890$        & $100$        & $310$        & $720$        & $150$         \\
        \midrule
        $\kappa_1$          & $0.083$                                 & $0.084$      & $0.966$      & $0.073$      & $0.070$      & $0.968$       \\
        $\kappa_2$          & $0.007$                                 & $0.976$      & $0.094$      & $0.223$      & $-0.966$     & $0.045$       \\
        $\kappa_3$          & $-0.997$                                & $-0.203$     & $0.239$      & $0.972$      & $-0.250$     & $0.246$       \\
        \bottomrule
    \end{tabularx}
    \caption{Input parameter values and coupling scale factors, $\kappa_a$
    ($a=1,2,3$), for the six defined benchmark scenarios. The doublet vev is set
    to $v=\SI{246}{\GeV}$ for all scenarios.}
    \label{tab:BPparams}
\end{table}

For this whitepaper, I rescanned all benchmark planes with the newest \HiggsBounds and \HiggsSignals versions: \HBv{5.10.2} and \HSv{2.6.2}. For nearly all parameter points, these new versions did not introduce additional constraints on the parameter space, and I therefore show the benchmark planes from the original publication. One exception is BP5 which has a slightly more constrained parameter space taking additional searches into account. I also comment on a possible recast on this plane and give a list of current experimental searches partially relying on our model.

\subsection{Asymmetric decays}
\begin{center}
\begin{figure} 
\begin{center}
\begin{minipage}{0.45\textwidth}
\begin{center}
\includegraphics[width=\textwidth]{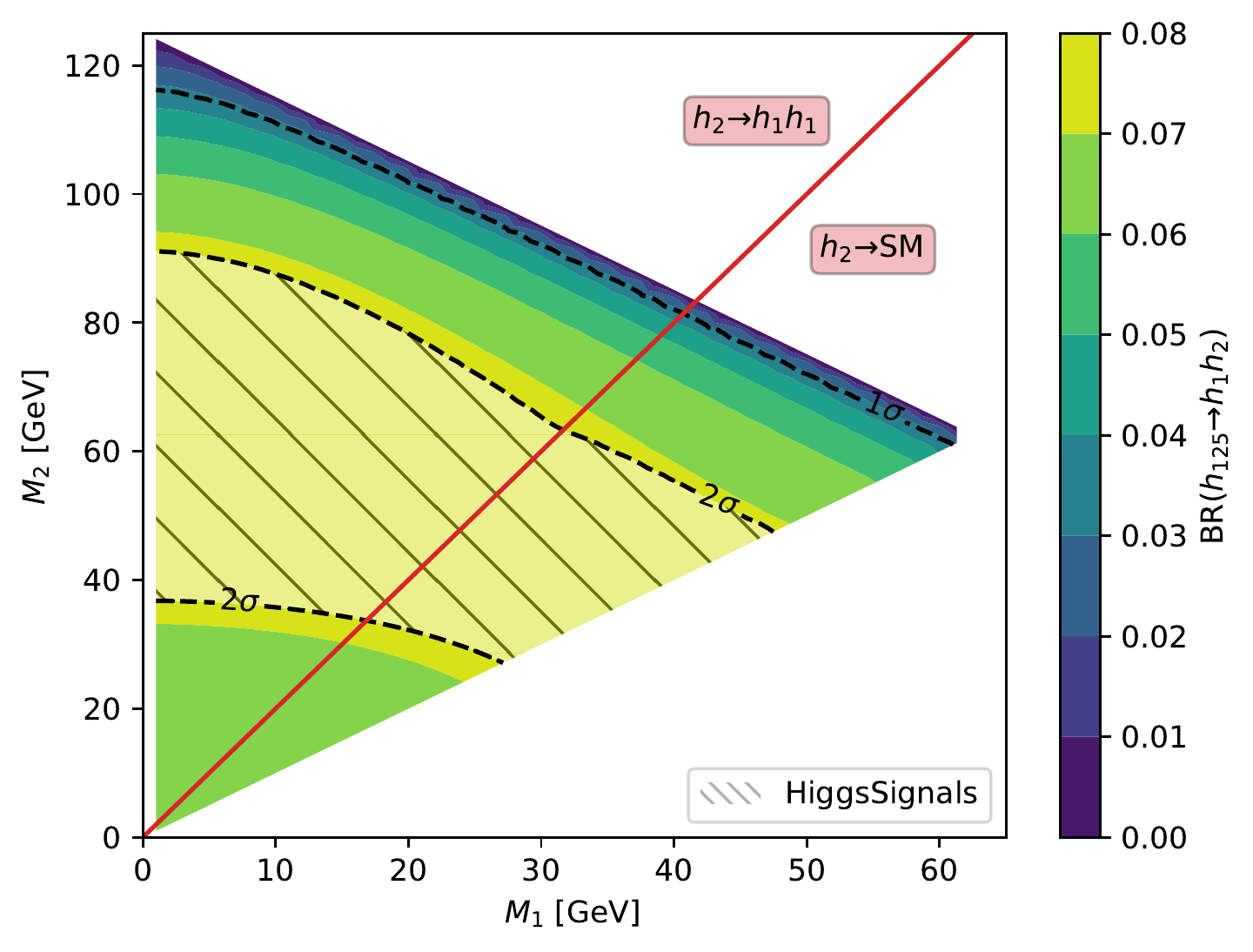}\\
{\small Benchmark Plane 1}
\end{center}
\end{minipage}\\ 
\vspace{6mm}
\begin{minipage}{\textwidth}
\begin{center}
\includegraphics[width=0.45\textwidth]{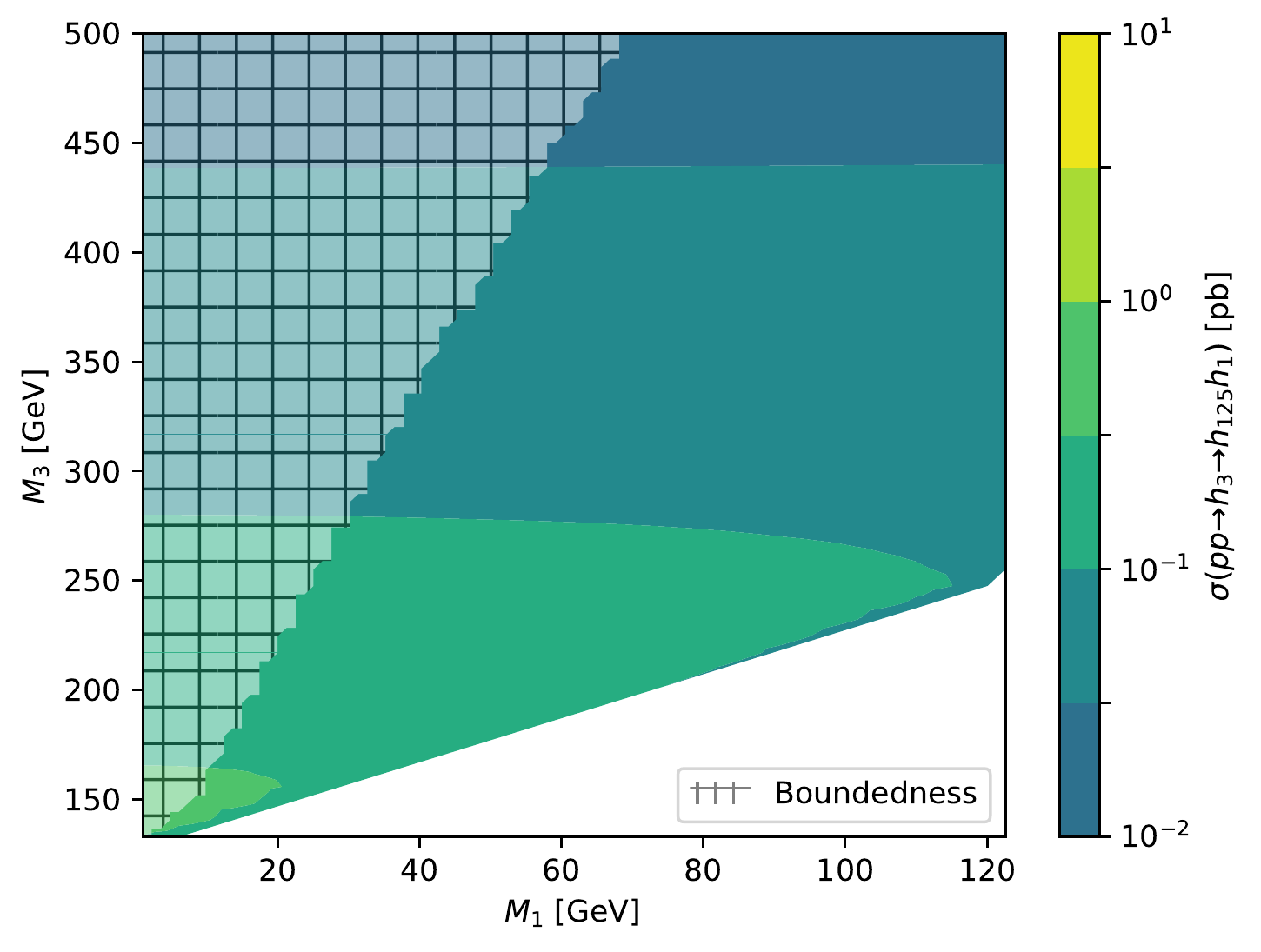}
\includegraphics[width=0.45\textwidth]{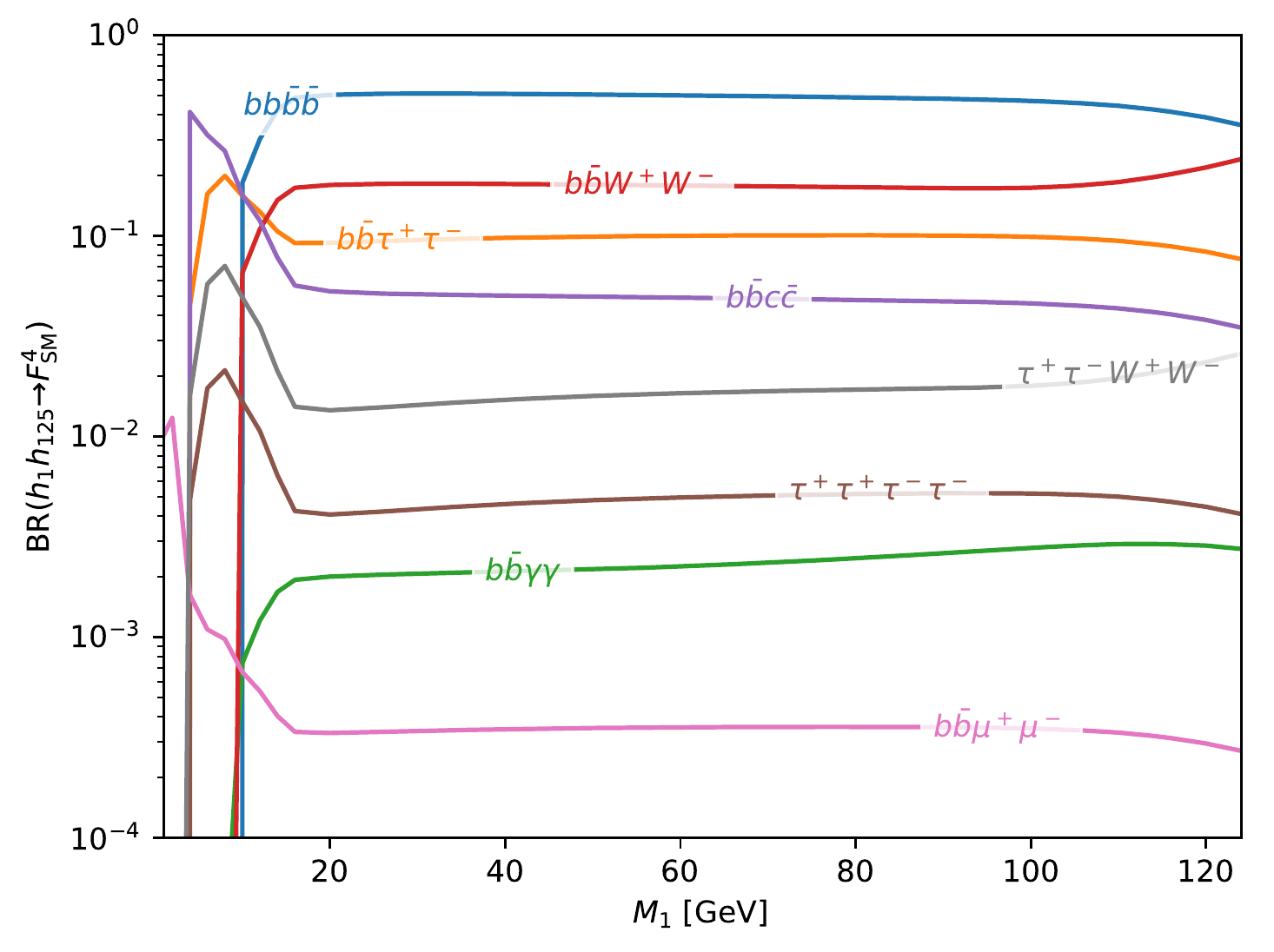}\\
{Benchmark Plane 2}
\end{center}
\end{minipage}\\
\vspace{6mm}
\begin{minipage}{\textwidth}
\begin{center}
\includegraphics[width=0.45\textwidth]{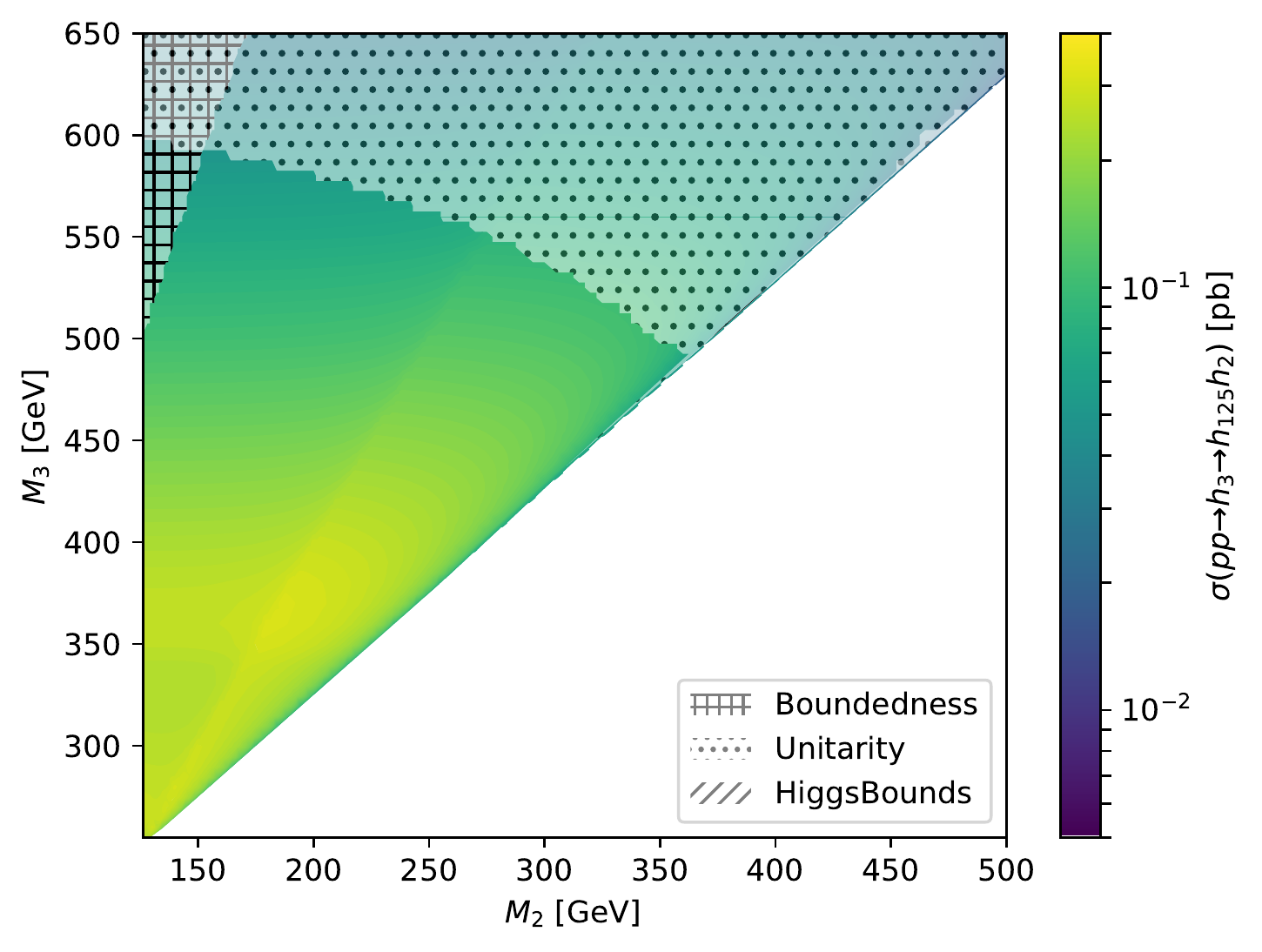}
\includegraphics[width=0.49\textwidth]{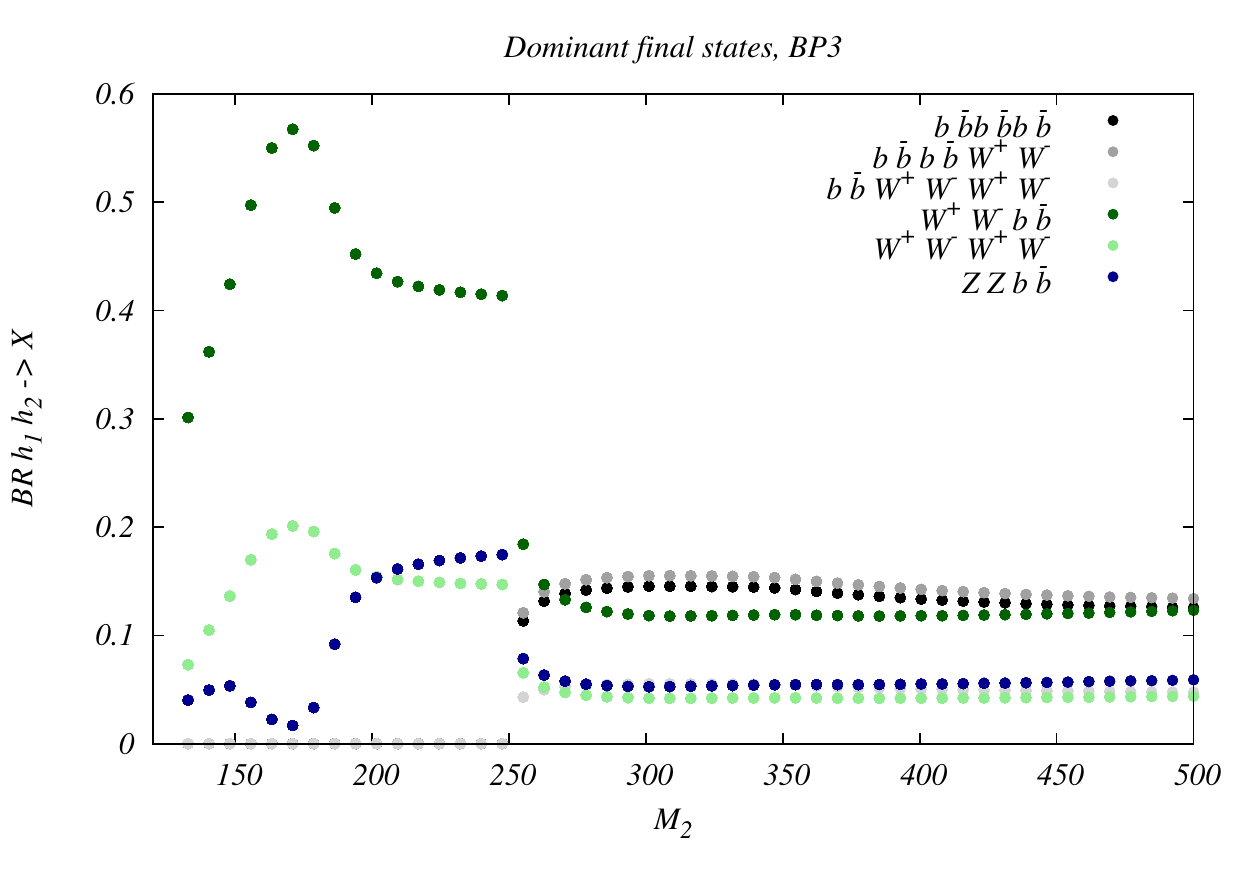}\\
{Benchmark Plane 3}
\end{center}
\end{minipage}\\
\end{center}
\caption{\label{fig:asymm} Benchmark planes for asymmetric production and decay, $p\,p\,\rightarrow\,h_{3}\,\rightarrow\,h_1\,h_2$, for various assigments of the 125 \GeV~ resonance. {\sl Top row:} BP1, where $h_3\,\equiv\,h_{125}$. Production cross sections are close to the SM production here, of around $\sim\,4\,\pb$ at 13 \TeV. Shown is the branching ratio to $h_1\,h_2$ in the two-dimensional mass plane. {\sl Middle and bottom rows:} BPs 2 and 3, where $h_{2,1}\,\equiv\,h_{125}$, respectively. {\sl Left:} Production cross sections at a 13 \TeV~ LHC. {\sl Right: } Branching ratios of the $h_1\,h_2$ state as a function of the free light scalar mass. Partially taken from \cite{Robens:2019kga}.}
\end{figure}
\end{center}

In this subsection, I discuss the asymmetric decay modes $h_3\,\rightarrow\,h_1\,h_2$, where successively one of the three scalars is identified with the 125 \GeV~ resonance. I display the corresponding benchmark planes in figure \ref{fig:asymm}.
Depending on the benchmark plane, maximal production cross sections are given by $\sim\,3-4\,\pb,\,\sim\,0.6\,\pb,$ and $0.3\,\pb$ for $h_1\,h_2$ production for BPs 1/2/3, respectively. In BP3, the $h_1h_1h_1$ final state reaches cross sections up to $\sim\,140\,\fb$. Note that as soon as the kinematic threshold for $h_2\,\rightarrow\,h_{125} h_{125}$ is reached, in fact decays from that state become dominant.
\subsection{Symmetric decays}
\begin{center}
\begin{figure} 
\begin{center}
\begin{minipage}{\textwidth}
\begin{center}
\includegraphics[width=0.45\textwidth]{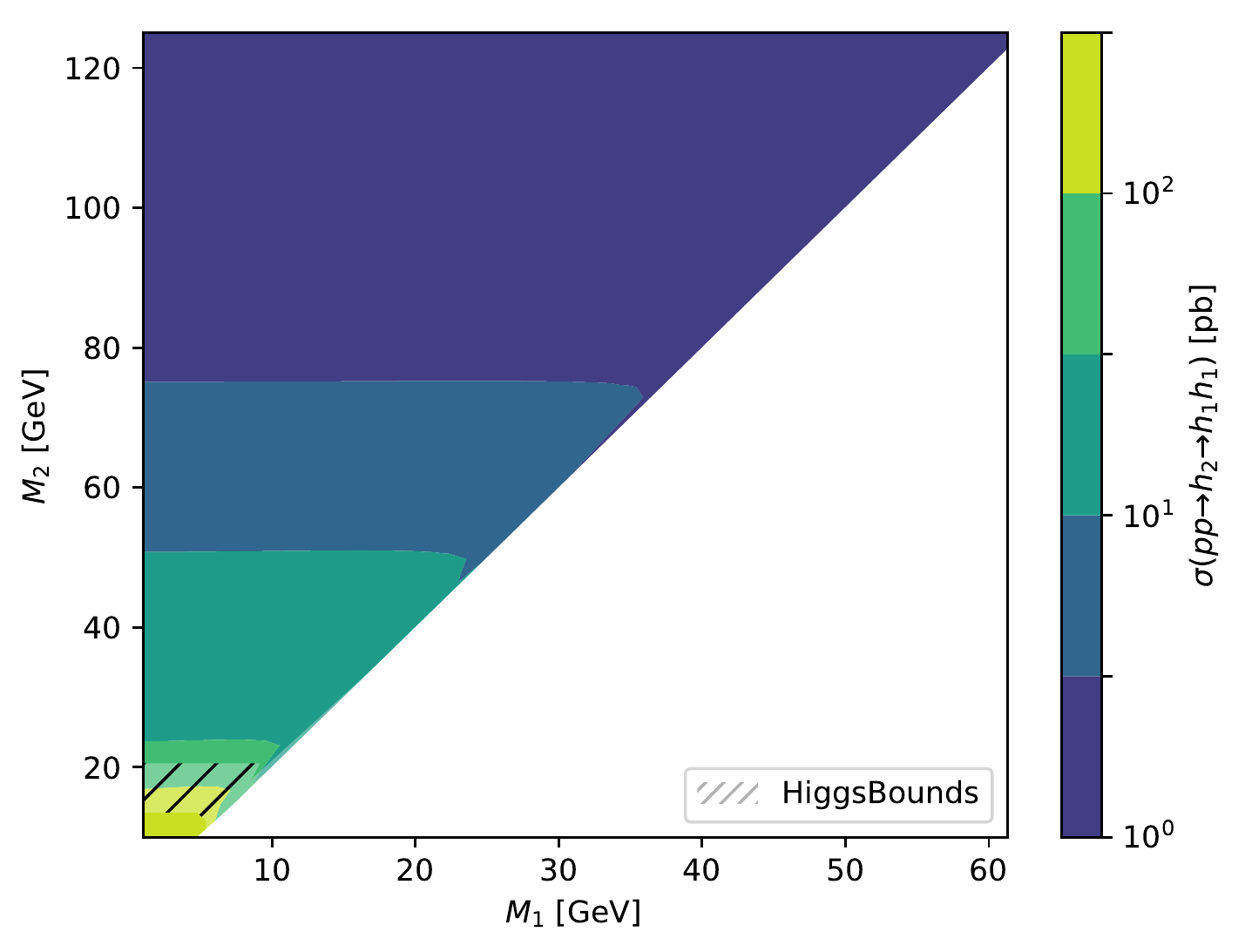}
\includegraphics[width=0.45\textwidth]{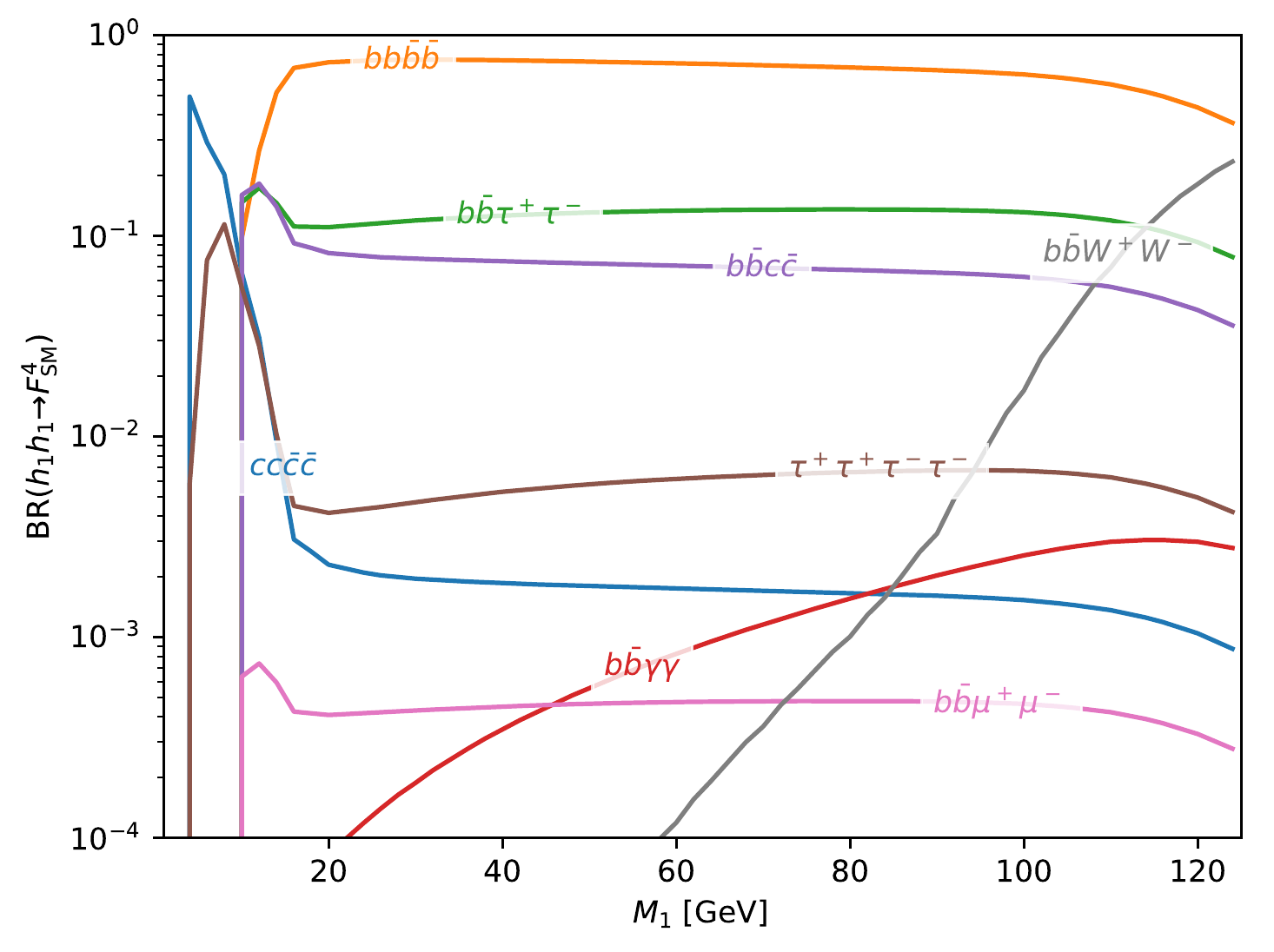}
\\
\vspace{-2mm}
{\small Benchmark Plane 4}
\end{center}
\end{minipage}\\ 
\vspace{6mm}
\begin{center}
\begin{minipage}{0.45\textwidth}
\begin{center}
\includegraphics[width=\textwidth]{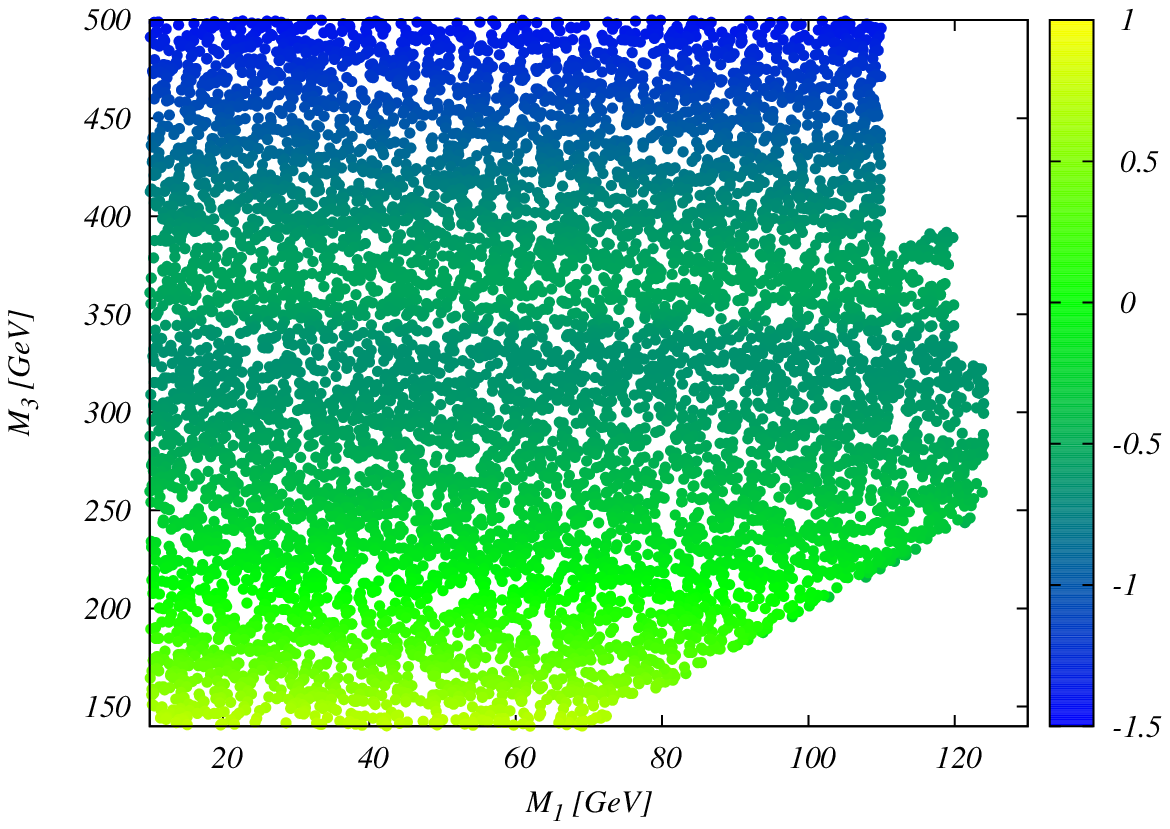}
\end{center}
\end{minipage}
\\
{\small Benchmark Plane 5 [color coding: $\log_{10}\lb \sigma_{h_3\,\rightarrow\,h_1\,h_1}/[\pb]\rb$]}
\end{center}
\vspace{6mm}
\begin{minipage}{\textwidth}
\begin{center}
\includegraphics[width=0.45\textwidth]{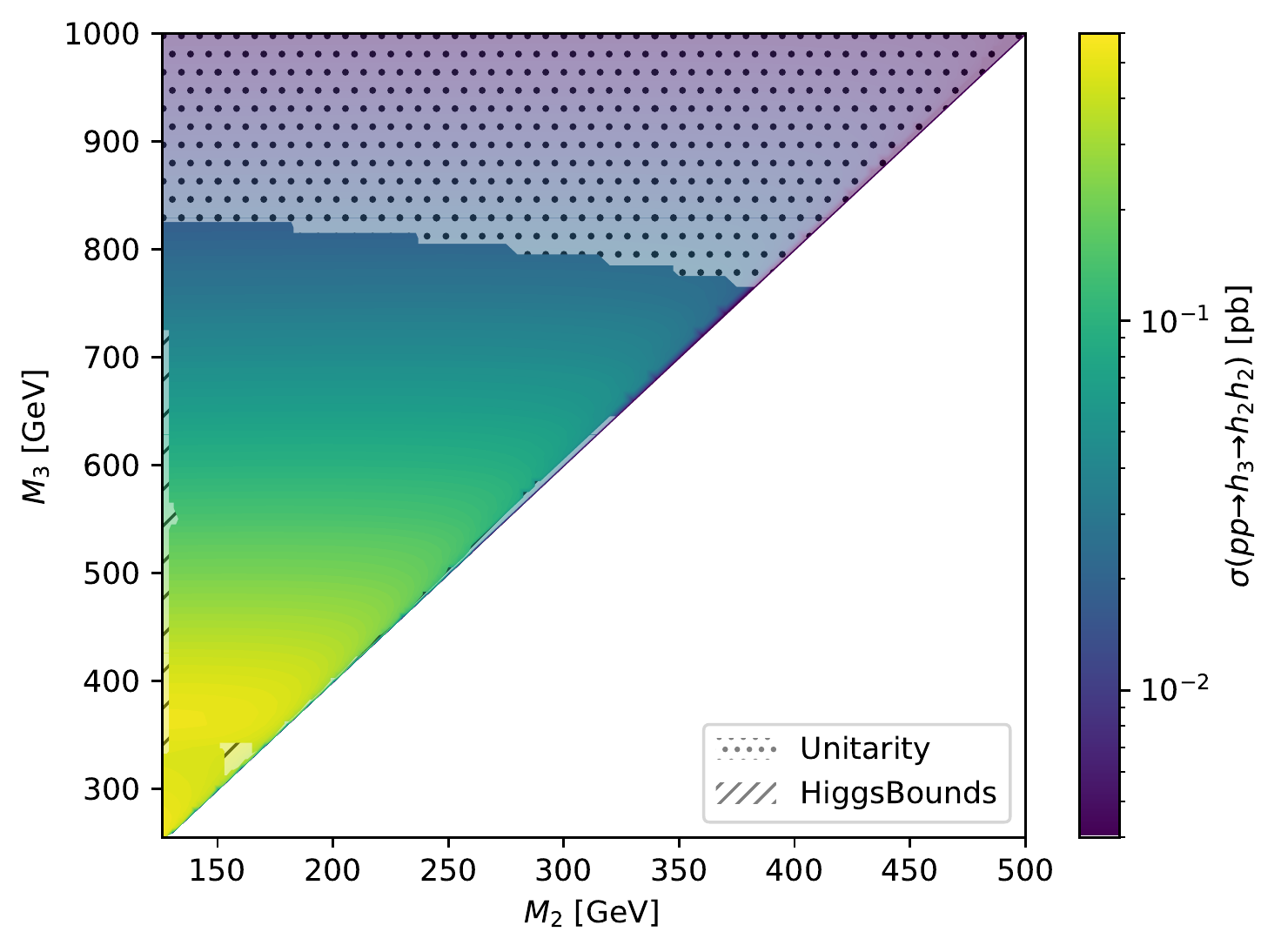}
\includegraphics[width=0.49\textwidth]{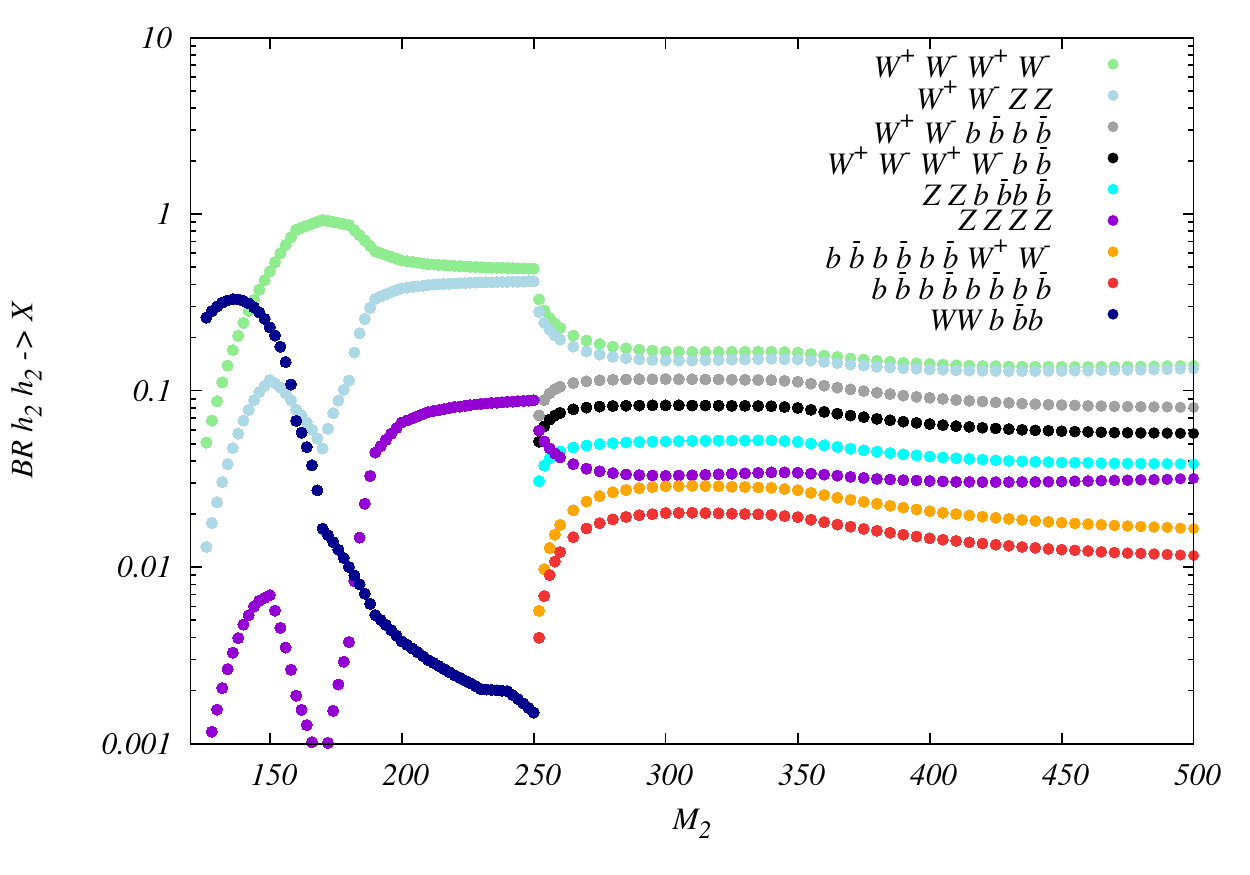}\\
\vspace{-2mm}
{Benchmark Plane 6}
\end{center}
\end{minipage}\\
\end{center}
\vspace{4mm}
\caption{\label{fig:symm}  Benchmark planes for symmetric production and decay, $p\,p\,\rightarrow\,h_{i}\,\rightarrow\,h_j\,h_j$, for various assigments of the 125 \GeV~ resonance. {\sl Top /middle/ bottom rows:} BPs 4/5/6, where $h_{3/2/1}\,\equiv\,h_{125}$. {\sl Left:} Production cross sections at a 13 \TeV~ LHC. {\sl Right: } Branching ratios of the $h_j\,h_j$ state as a function of the lighter free scalar mass. Branching ratios for BP4 and 5 are identical, therefore only one plot is displayed here. Partially taken from \cite{Robens:2019kga}.}
\end{figure}
\end{center}

Symmetric decays are given by BPs 4/5/6, with again a differing assignment for $h_{3/2/1}\,\equiv\,h_{125}$, respectively. The corresponding production and decay modes are displayed in figure \ref{fig:symm}.
Depending on the benchmark plane, pair-production cross sections can reach up to 
60/ 2.5/ 0.5 \pb~ for BPs 4/5/6, respectively. For the latter the $h_{125}h_{125} h_{125} h_{125}$ final state can reach rates up to 14 \fb. Also note that the allowed parameter space in BP5 has slightly shrunk, mainly due to the implementation of an additional search \cite{ATLAS:2018rnh} into HiggsBounds after the performance of the original scan. For BP6, 6 particle final states as e.g. $W^+W^-b\bar{b}b\bar{b}$ can reach branching ratios up to $\sim\,10\%$, depending on $M_2$.
\section{Further investigation of this model}
After the original appearance of the paper proposing the TRSM, several theoretical and experimental works have been performed which at least partially build on the benchmark planes proposed here. We briefly list some of these here.
\subsection{Investigation of the $h_{125}h_{125}h_{125}$ final state}
In BP3, for $M_2\,\rightarrow\, 250\,\GeV$, the decay $h_{2}\,\rightarrow\,h_1\,h_1$ becomes dominant, leading to a $h_{125}h_{125}h_{125}$ final state. For subsequent decays into $b\,\bar{b}$, this BP has been investigated in \cite{Papaefstathiou:2020lyp}. We found that, depending on the parameter point and integrated luminosity, significances between 3 and $\sim\,10$ can be achieved. We display the results in table \ref{table:efficiencies}

\begin{center}
\begin{figure} 
\begin{center}
  \includegraphics[width=0.6\columnwidth]{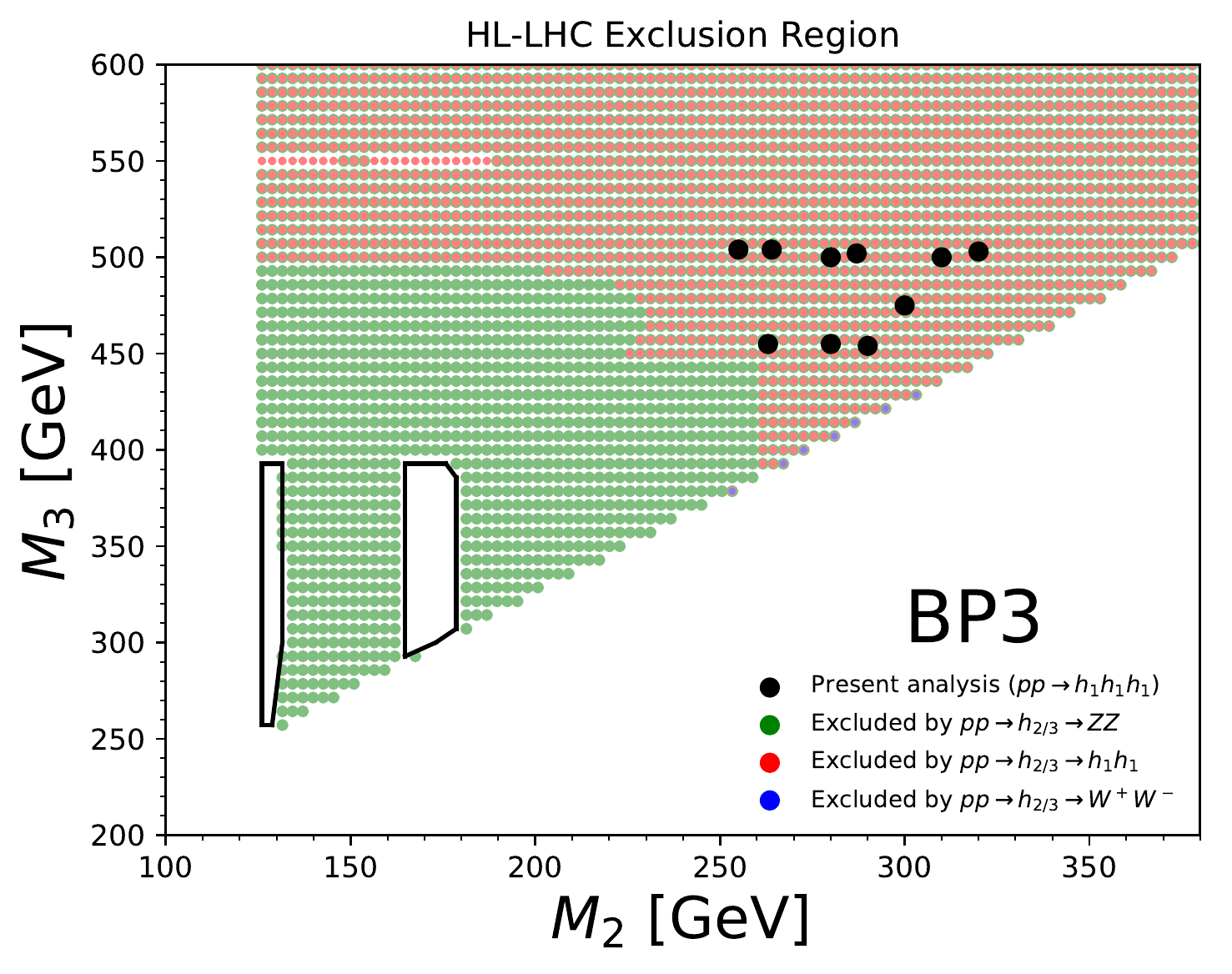}
\caption{\label{fig:const_Andreas} The expected exclusion region for the full integrated luminosity of the HL-LHC, $3000~\rm{fb}^{-1}$, through final states \textit{other} than $p p \rightarrow h_1 h_1 h_1$ as explained in the main text. Points with green circles are expected to be excluded by $ZZ$ final states, with red circles by $h_1 h_1$ and with blue circles by $W^+W^-$. The $W^+W^-$ analysis excludes only very few points on the parameter space and therefore appears infrequently in the figure. The points \textbf{A}--\textbf{I} that we have considered in our analysis of $p p \rightarrow h_1 h_1 h_1$ are shown in black circles overlayed on top of the circles indicating the exclusion. The two cut-out white regions near $M_2 \sim 130$~GeV and $M_2 \sim 170$~GeV will remain viable at the end of the HL-LHC. Taken from \cite{Papaefstathiou:2020lyp}.}
\end{center}
\end{figure}
\end{center}

\begin{table*}[t!]
\begin{tabular*}{\textwidth}{@{\extracolsep{\fill}}cccccccc@{}}
Label&$(M_2, M_3)$ & $\varepsilon_{\rm Sig.}$& $\rm{S}\bigl|_{300\rm{fb}^{-1}}$ & $\varepsilon_{\rm Bkg.}$ & 
$\rm{B}\bigl|_{300\rm{fb}^{-1}}$ & $\text{sig}|_{300\rm{fb}^{-1}}$ & $\text{sig}|_{3000\rm{fb}^{-1}}$\\
& [GeV] & & & & & (syst.) &(syst.)  
\\
\hline
\textbf{A} &$(255, 504)$ & $0.025$ & $14.12$  & $8.50\times 10^{-4}$ & $19.16$ & $2.92~(2.63)$&$9.23~(5.07)$\\
\textbf{B} & $(263, 455)$ & $0.019$ & $17.03$    & $3.60\times 10^{-5}$ & $ {8.12}$ & $4.78 ~(4.50)$&$15.10~(10.14)$\\
\textbf{C} & $(287, 502)$ & $0.030$ & $20.71$ & $9.13\times 10^{-5}$ & $20.60$  & $4.01~(3.56)$ & $12.68~(6.67)$\\
\textbf{D} & $(290, 454)$ & $0.044$ & $37.32$    & $1.96\times 10^{-4}$ & $44.19$& $5.02~(4.03)$&$15.86~(6.25)$\\
\textbf{E} & $(320, 503)$ & $0.051$ & $ {31.74}$    & $2.73\times 10^{-4}$ & $61.55$& $3.76~( {2.87}) $&$11.88~(4.18)$\\
\textbf{F} & $(264,504)$&$0.028$& $18.18$&$9.13\times 10^{-5}$&$20.60$&$3.56~(3.18) $&$11.27~(5.98)$\\
\textbf{G} & $(280, 455)$&$0.044$& $38.70$ &$1.96\times 10^{-4}$& $44.19$ & $5.18~(4.16)$ &$16.39~(6.45)$\\
\textbf{H} & $(300, 475)$ & $0.054$& $41.27$ & $2.95\times 10^{-4}$& $66.46$ & $4.64~(3.47)$&$ 14.68~( {4.94})$\\
\textbf{I} & $(310, 500)$& $0.063$& $41.43$& $3.97\times 10^{-4}$& $89.59$& $4.09~(2.88) $&$ {12.94~(3.87)}$\\
\textbf{J} & $(280,500)$& $0.029 $& $20.67$&$9.14\times 10^{-5}$& $20.60$&$4.00~(3.56) $&$12.65~(6.66)$\\
\end{tabular*}
\caption{ The resulting selection efficiencies, $\varepsilon_{\rm Sig.}$ and $\varepsilon_{\rm Bkg.}$, number of events, $S$ and $B$ for the signal and background, respectively, and statistical significances. A $b$-tagging efficiency of $0.7$ has been assumed. The number of signal and background events are provided at an integrated luminosity of $300~\rm{fb}^{-1}$. Results for $3000~\rm{fb}^{-1}$ are obtained via simple extrapolation. The significance is given at both values of the integrated luminosity excluding (including) systematic errors in the background. Taken from \cite{Papaefstathiou:2020lyp}.}
\label{table:efficiencies}
\end{table*}
Note we also compared how different channels, e.g. direct decays of the heavier scalars into $VV$ or $h_{125}h_{125}$ final states, would perform at a HL-LHC. The results are displayed in figure \ref{fig:const_Andreas}.
We note that all benchmark points that were investigated can additionally be probed by other production and decay mechanisms. Note, however, that these test different regions of the parameter space, as they depend on different parameters in the potential. These searches can therefore be considered to be complementary.

\subsection{Recasting current LHC searches}

\begin{figure} 
\begin{center}
\begin{minipage}{0.45\textwidth}
\includegraphics[width=\textwidth]{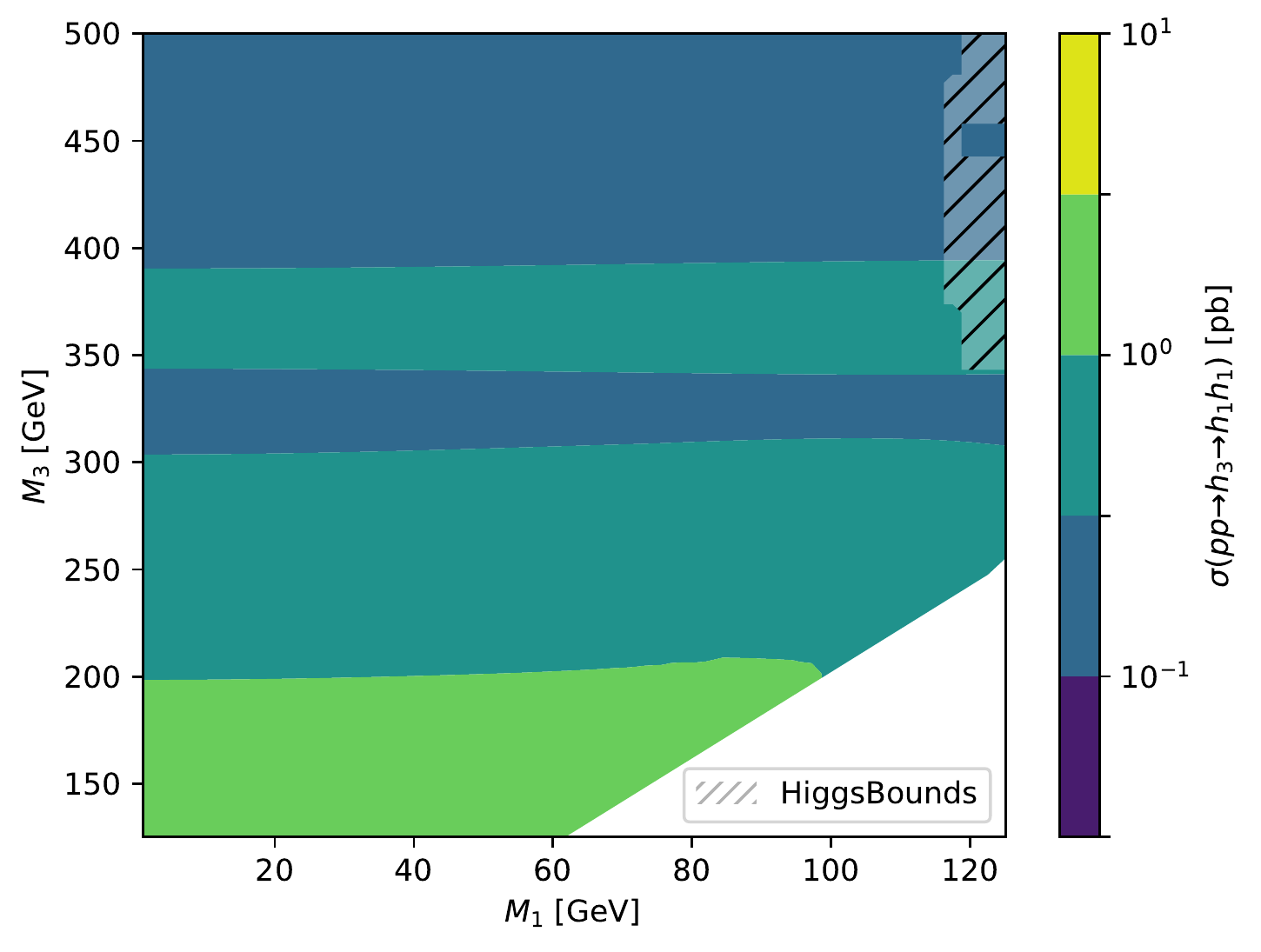}
\end{minipage}
\begin{minipage}{0.45\textwidth}
\includegraphics[width=\textwidth]{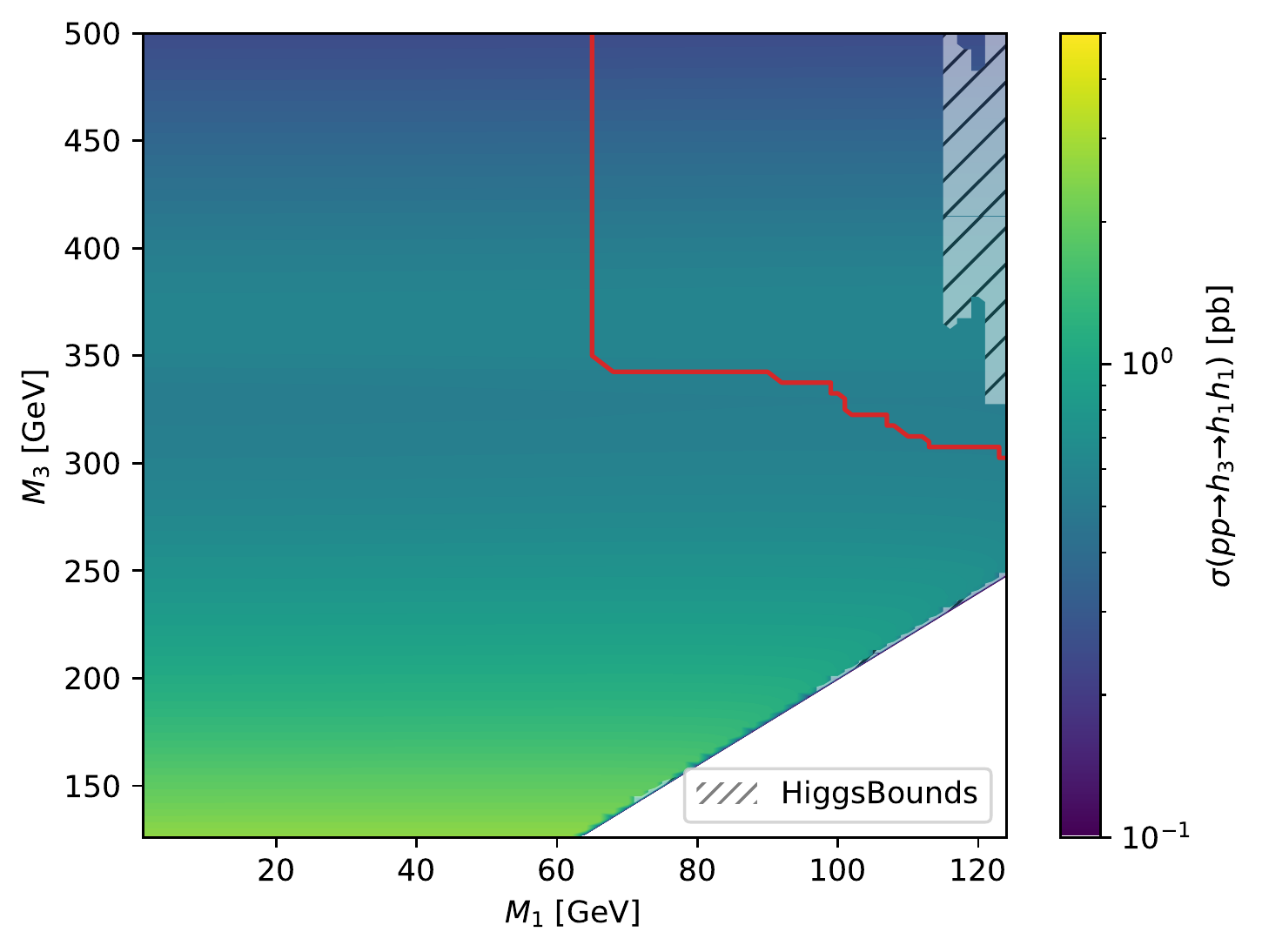}
\end{minipage}
\end{center}
\caption{\label{fig:bp5reint} Reinterpretation of a $36\,\fb^{-1}$ CMS search for di-Higgs production via a heavy resonance using the 4 b final state. The exclusion line uses the results obtained in \cite{Barducci:2019xkq}. Points to the right and above the red contour are excluded. Taken from \cite{Robens:2022mvi}.}
\end{figure}

It is also interesting to investigate whether current searches can be reinterpreted and recasted in such a way that they allow to exclude regions in the models parameter space that were not directly scrutinized in the experimental search, or for which no interpretation was presented in the original publication. In \cite{Barducci:2019xkq}, the authors have reinterpreted a CMS search for $p\,p\,\rightarrow\,H\,\rightarrow\,h_{125}h_{125}\,\rightarrow\,4\,b$ \cite{CMS:2018qmt}, which corresponds to di-Higgs production via a heavy resonance and subsequent decays into $b\,\bar{b}$ final states, and extended the mass ranges for the scalars in the decay chain. I have applied these results to the TRSM, in particular to BP5. I display the corresponding results in figure \ref{fig:bp5reint}\footnote{I thank the authors of \cite{Barducci:2019xkq} for providing us with the corresponding exclusion limits.}. We see that the sensitive region of parameter space is significantly extended, and therefore, an actual experimental analysis also in this parameter region is greatly encouraged.
\subsection{Experimental searches with TRSM interpretations}
\begin{center}
\begin{figure} 
\includegraphics[width=\textwidth]{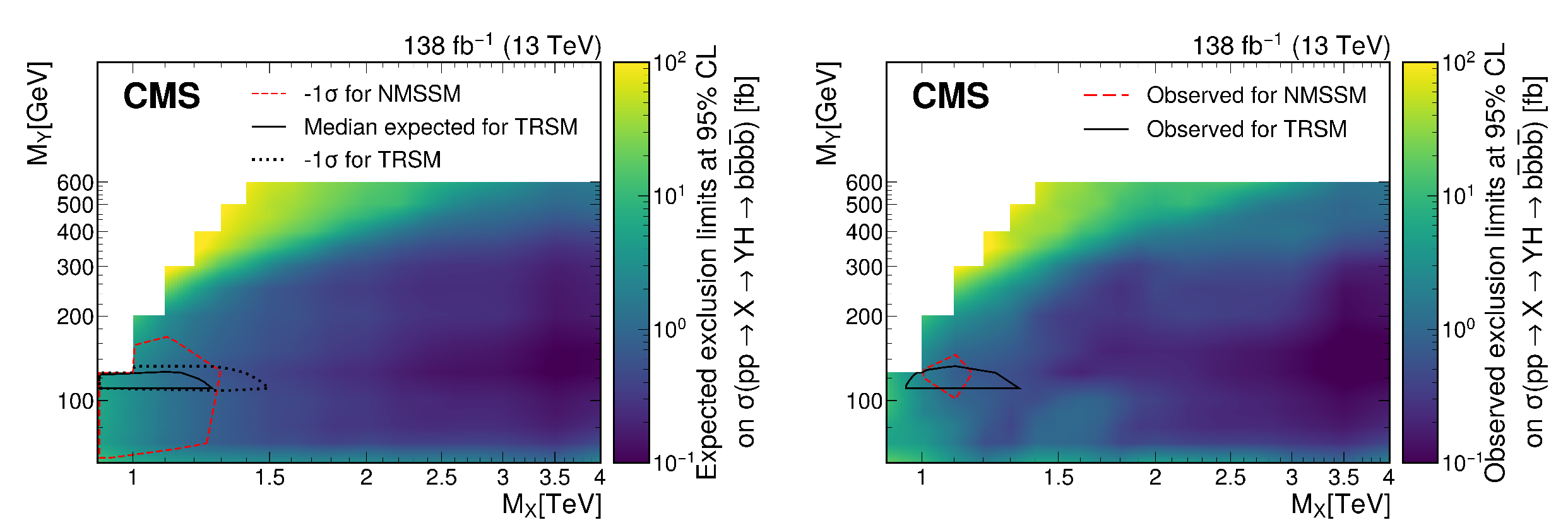}
\caption{\label{fig:cmsres} Expected {\sl (left)} and observed {\sl (right)} $95\%$ confidence limits for the $p\,p\,\rightarrow\,h_3\,\rightarrow\,h_2\,h_1$ search, with subsequent decays into $b\bar{b}b\bar{b}$. For both models, maximal mass regions up to $m_3\,\sim\,1.4\TeV,\;m_2\,\sim\,140\,\GeV$ can be excluded. Figure taken from \cite{CMS:2022suh}.}
\end{figure}
\end{center}

At least one experimental search actually has made use of the predictions obtained within the TRSM to interpret regions in parameter space that are excluded: a CMS search for asymmetric production and subsequent decay into $b\bar{b}b\bar{b}$  final states \cite{CMS:2022suh}. For this, maximal production cross sections were provided in the parameter space, allowing all additional new physics parameter to float; the respective values have been tabulated in \cite{reptr}. Figure \ref{fig:cmsres} shows the expected and observed limits in this search for the TRSM and NMSSM \cite{Ellwanger:2022jtd}.
In addition, several searches also investigate decay chains that can in principle also be realized within the TRSM, as e.g. other searches for the same final states \cite{CMS:2021erw,CMS:2021fjj,CMS:2022qww} or $b\,\bar{b}\mu^+\mu^-$ \cite{ATLAS:2021ypo,ATLAS:2021hbr} final states.

\section{Signatures at Higgs factories}
\begin{center}
\begin{figure} 
\begin{center}
\includegraphics[width=0.45\textwidth]{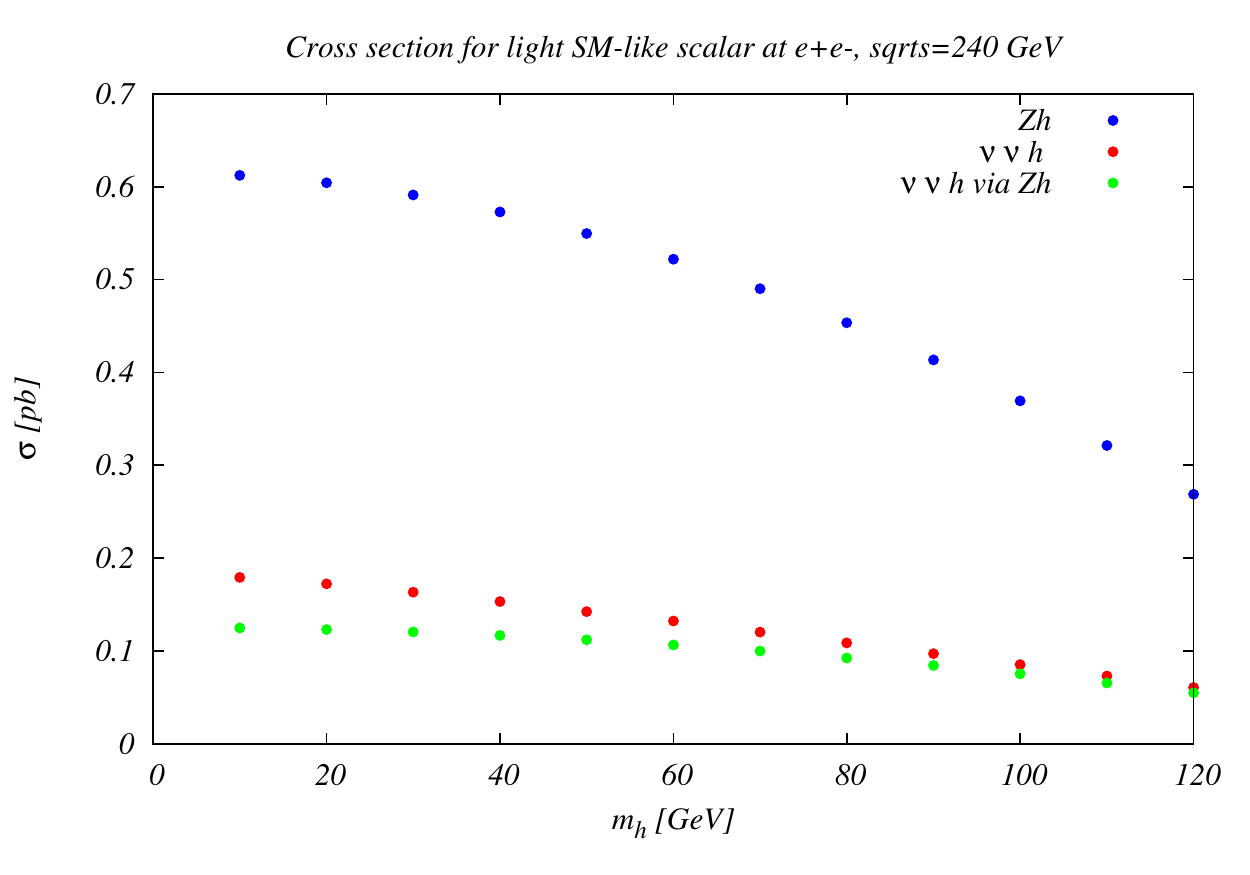}
\includegraphics[width=0.45\textwidth]{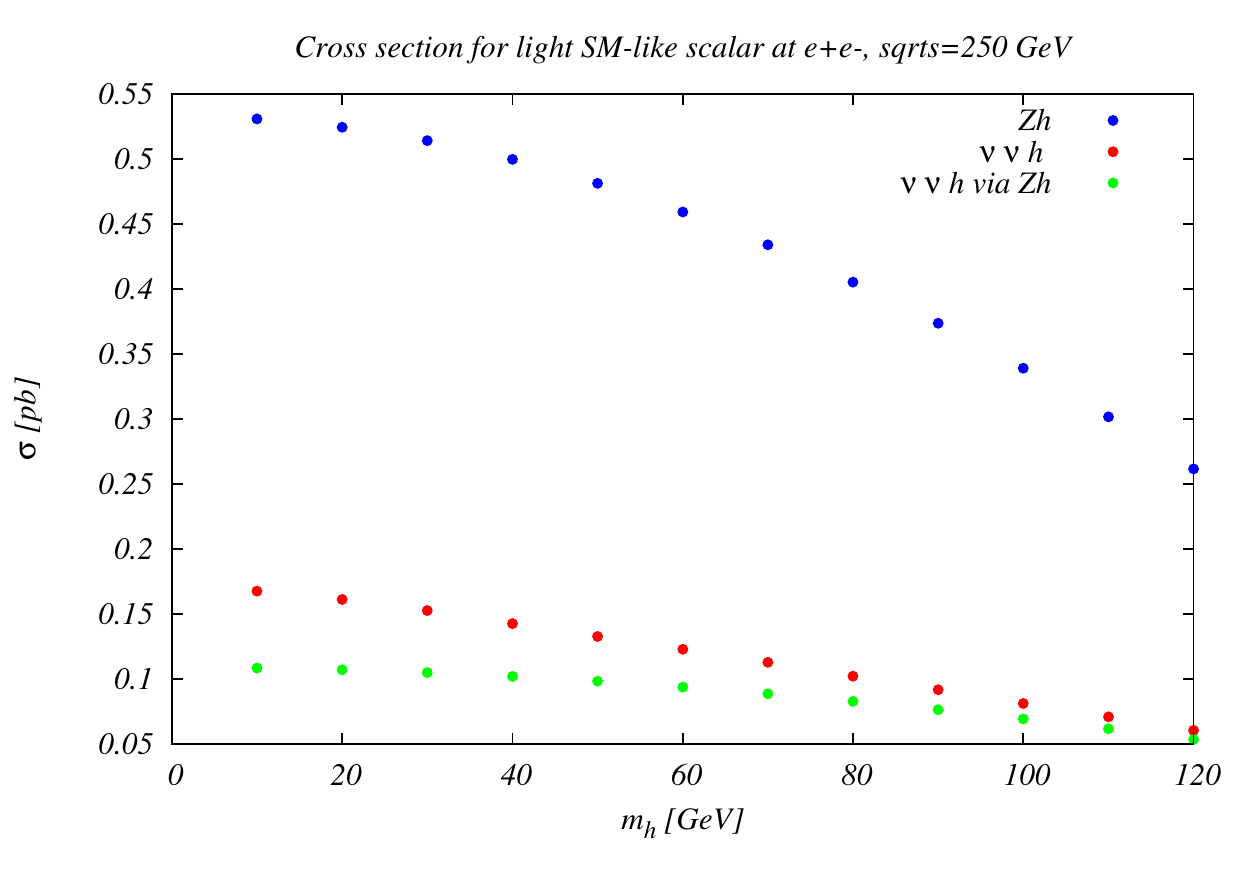}
\caption{\label{fig:prod250} Leading order production cross sections for $Z\,h$ and $h\,\nu_\ell\,\bar{\nu}_\ell$ production at an $e^+\,e^-$ collider with a com energy of 240 \GeV {\sl (left)} and 250 \GeV~ {\sl (right)} using Madgraph5 for an SM-like scalar h. Shown is also the contribution of $Z\,h$ to $\nu_\ell\,\bar{\nu}_\ell\,h$ using a factorized approach for the Z decay. Taken from \cite{Robens:2022zgk}.}
\end{center}
\end{figure}
\end{center}

The investigation of light scalars has recently gained again more interest, after the recommendation of the European Strategy Report \cite{EuropeanStrategyforParticlePhysicsPreparatoryGroup:2019qin,European:2720129} to concentrate on $e^+e^-$ machines with $\sqrt{s}\,\sim\,240-250\,\GeV$. A short review about the current state of the art for such searches and models which allow for low scalars can e.g. be found in \cite{Robens:2022zgk}. In this model, the only feasible production is $Zh$ radiation of the lighter scalar, with production cross sections given in figure \ref{fig:prod250}. Cross sections have been derived using Madgraph5 \cite{Alwall:2011uj}.

We can now investigate what would be production cross sections for scalar particles with masses $\lesssim\,160\,\GeV$ at Higgs factories.
\subsection{Production of 125 \GeV~ resonance and subsequent decays}
We first turn to the easy case of the production of the 125 \GeV~ resonance in various benchmark scenarios. Of interest are cases where decays $h_{125}\,\rightarrow\,h_i\,h_j$ are kinematically allowed. Note that our benchmark points were not set up in particular for the scenario where $i\,=\,j$, so for this rates might be relatively small by construction.

From table \ref{tab:BPparams} we see that for all scenarios the rescaling for the 125 \GeV~ resonance is $\gtrsim\,0.966$, leading to production cross sections of about $\sim\,0.2\,\pb$, close to the SM value. In general, due to constraints from the invisible branching ratio \cite{ATLAS-CONF-2020-052} as well as signal strength fits, the production cross section for $h_i\,h_j$ final states has to be lower by at least an order of magnitude, leading to cross sections $\mathcal{O}\lb 10\,\fb\rb$. In fact, in the benchmark planes presented here the largest rate for $Z\,h_{125}$ production and subsequent scalar decays can be found in BP1, where the rates are given by multiplying the BRs from figure \ref{fig:asymm} with the production of $Z\,h_{125}$, giving maximal cross sections of around 18 \fb.

\subsection{Additional scalar production}
\begin{center}
\begin{figure} 
\begin{center}
\includegraphics[width=0.45\textwidth]{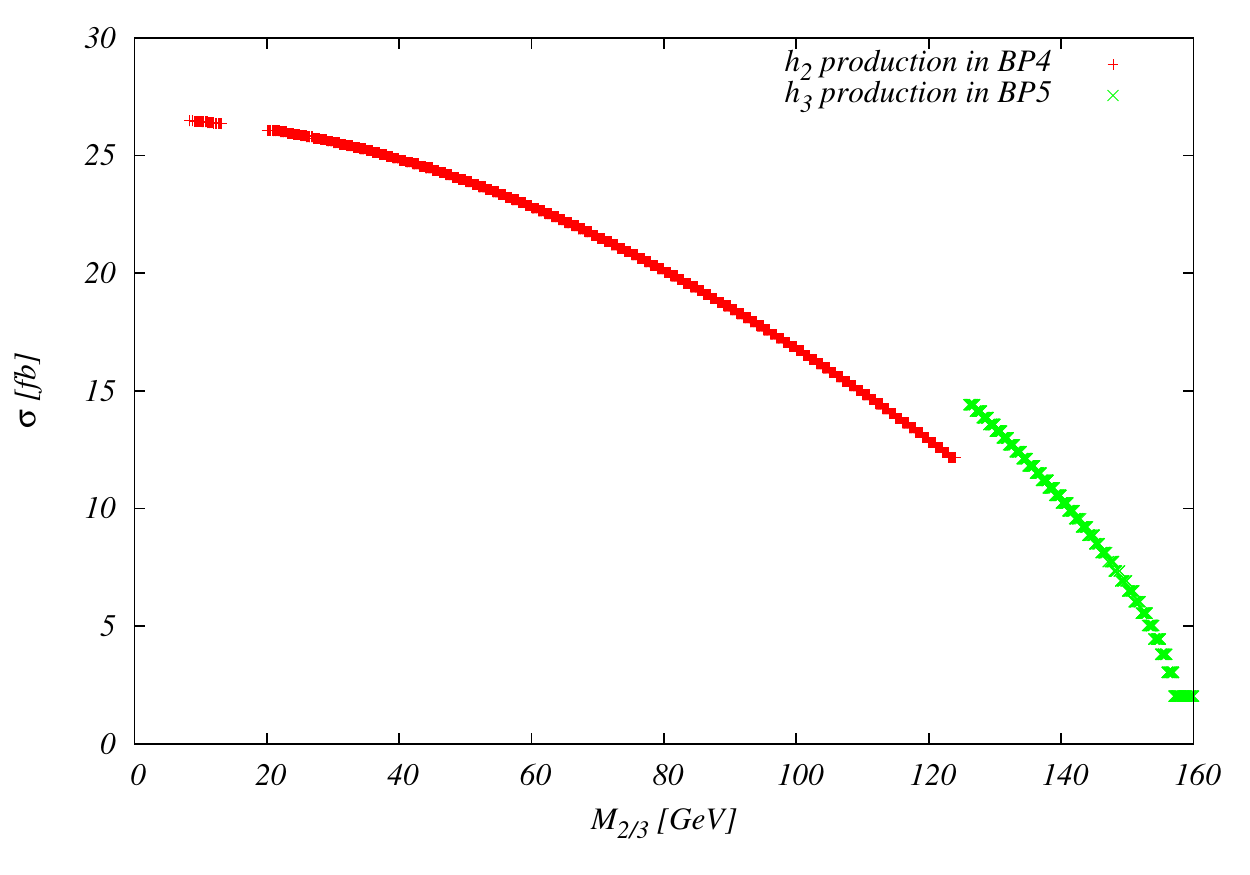}
\end{center}
\caption{\label{fig:prod45} Production cross sections for $Z h_{2/3}$ in BPs 4 and 5, respectively, at a 250 \GeV Higgs factory.}
\end{figure}
\end{center}

We now turn to the Higgs-Strahlung production of new physics scalars. This process is in principle possible in all BPs discussed here. However, if we require production rates of $Z\,h_i$ to be larger than $\sim\,10\,\fb$, only BPs 4 and 5 render sufficiently large rates for the production of $h_2$ and $h_3$, respectively. Production rates are independent of the other scalars, and we therefore depict them for both BPs in figure \ref{fig:prod45}
BP4 is constructed in such a way that as soon as the corresponding parameter space opens up, the $h_1\,h_1$ decay becomes dominant; final states are therefore mainly $Z\,b\bar{b}b\bar{b}$ if $M_2\,\gtrsim\,2\,M_1$. Below that threshold, dominant decays are into a $b\,\bar{b}$ pair, which means that standard searches as e.g. presented in \cite{Drechsel:2018mgd,Wang:2020lkq} should be able to cover the parameter space.

Similarly, in BP5 the $h_3\,\rightarrow\,h_1\,h_1$ decay is also favoured as soon as it is kinematically allowed. Therefore, in this parameter space again $Z b\bar{b}b\bar{b}$ final states become dominant. Otherwise $Z\,b\bar{b}$ and $ZW^+W^-$ final states prevail, with a cross over for the respective final states at around $M_3\,\sim\,135\,\GeV$. Branching ratios for these final states are in the $40-50\%$ regime.

\section{Summary}
In this whitepaper, I gave a short summary of the status of collider signatures and searches in the TRSM introduced in \cite{Robens:2019kga}. I gave a summary on current state of the art and investigation, including further detailed collider studies, recasts, as well as current searches that use or are motivated in this model. I also gave a brief overview on channels within this model that might be testable at future $e^+e^-$ machines, with a focus in Higgs factories with $\sqrt{s}\,\sim\,250\,\GeV$.

\bibliography{lit}

\begin{thebibliography}{10}

\bibitem{Robens:2019kga}
Tania Robens, Tim Stefaniak, and Jonas Wittbrodt.
\newblock {Two-real-scalar-singlet extension of the SM: LHC phenomenology and
  benchmark scenarios}.
\newblock {\em Eur. Phys. J. C}, 80(2):151, 2020, 1908.08554.

\bibitem{Coimbra:2013qq}
Rita Coimbra, Marco O.~P. Sampaio, and Rui Santos.
\newblock {ScannerS: Constraining the phase diagram of a complex scalar singlet
  at the LHC}.
\newblock {\em Eur. Phys. J.}, C73:2428, 2013, 1301.2599.

\bibitem{Ferreira:2014dya}
P.~M. Ferreira, Renato Guedes, Marco O.~P. Sampaio, and Rui Santos.
\newblock {Wrong sign and symmetric limits and non-decoupling in 2HDMs}.
\newblock {\em JHEP}, 12:067, 2014, 1409.6723.

\bibitem{Costa:2015llh}
Raul Costa, Margarete M\"uhlleitner, Marco O.~P. Sampaio, and Rui Santos.
\newblock {Singlet Extensions of the Standard Model at LHC Run 2: Benchmarks
  and Comparison with the NMSSM}.
\newblock {\em JHEP}, 06:034, 2016, 1512.05355.

\bibitem{Muhlleitner:2016mzt}
Margarete M{\"{u}}hlleitner, Marco O.~P. Sampaio, Rui Santos, and Jonas
  Wittbrodt.
\newblock {The N2HDM under Theoretical and Experimental Scrutiny}.
\newblock {\em JHEP}, 03:094, 2017, 1612.01309.

\bibitem{Muhlleitner:2020wwk}
Margarete M\"uhlleitner, Marco O.~P. Sampaio, Rui Santos, and Jonas Wittbrodt.
\newblock {ScannerS: parameter scans in extended scalar sectors}.
\newblock {\em Eur. Phys. J. C}, 82(3):198, 2022, 2007.02985.

\bibitem{Bechtle:2008jh}
Philip Bechtle, Oliver Brein, Sven Heinemeyer, Georg Weiglein, and Karina~E.
  Williams.
\newblock {HiggsBounds: Confronting Arbitrary Higgs Sectors with Exclusion
  Bounds from LEP and the Tevatron}.
\newblock {\em Comput. Phys. Commun.}, 181:138--167, 2010, 0811.4169.

\bibitem{Bechtle:2011sb}
Philip Bechtle, Oliver Brein, Sven Heinemeyer, Georg Weiglein, and Karina~E.
  Williams.
\newblock {HiggsBounds 2.0.0: Confronting Neutral and Charged Higgs Sector
  Predictions with Exclusion Bounds from LEP and the Tevatron}.
\newblock {\em Comput. Phys. Commun.}, 182:2605--2631, 2011, 1102.1898.

\bibitem{Bechtle:2013gu}
Philip Bechtle, Oliver Brein, Sven Heinemeyer, Oscar Stal, Tim Stefaniak, Georg
  Weiglein, and Karina Williams.
\newblock {Recent Developments in HiggsBounds and a Preview of HiggsSignals}.
\newblock {\em PoS}, CHARGED2012:024, 2012, 1301.2345.

\bibitem{Bechtle:2013wla}
Philip Bechtle, Oliver Brein, Sven Heinemeyer, Oscar Stål, Tim Stefaniak,
  Georg Weiglein, and Karina~E. Williams.
\newblock {$\mathsf{HiggsBounds}-4$: Improved Tests of Extended Higgs Sectors
  against Exclusion Bounds from LEP, the Tevatron and the LHC}.
\newblock {\em Eur. Phys. J. C}, 74(3):2693, 2014, 1311.0055.

\bibitem{Bechtle:2015pma}
Philip Bechtle, Sven Heinemeyer, Oscar Stal, Tim Stefaniak, and Georg Weiglein.
\newblock {Applying Exclusion Likelihoods from LHC Searches to Extended Higgs
  Sectors}.
\newblock {\em Eur. Phys. J. C}, 75(9):421, 2015, 1507.06706.

\bibitem{Bechtle:2020pkv}
Philip Bechtle, Daniel Dercks, Sven Heinemeyer, Tobias Klingl, Tim Stefaniak,
  Georg Weiglein, and Jonas Wittbrodt.
\newblock {HiggsBounds-5: Testing Higgs Sectors in the LHC 13 TeV Era}.
\newblock {\em Eur. Phys. J. C}, 80(12):1211, 2020, 2006.06007.

\bibitem{Stal:2013hwa}
Oscar Stål and Tim Stefaniak.
\newblock {Constraining extended Higgs sectors with HiggsSignals}.
\newblock {\em PoS}, EPS-HEP2013:314, 2013, 1310.4039.

\bibitem{Bechtle:2013xfa}
Philip Bechtle, Sven Heinemeyer, Oscar Stål, Tim Stefaniak, and Georg
  Weiglein.
\newblock {$HiggsSignals$: Confronting arbitrary Higgs sectors with
  measurements at the Tevatron and the LHC}.
\newblock {\em Eur. Phys. J. C}, 74(2):2711, 2014, 1305.1933.

\bibitem{Bechtle:2014ewa}
Philip Bechtle, Sven Heinemeyer, Oscar Stål, Tim Stefaniak, and Georg
  Weiglein.
\newblock {Probing the Standard Model with Higgs signal rates from the
  Tevatron, the LHC and a future ILC}.
\newblock {\em JHEP}, 11:039, 2014, 1403.1582.

\bibitem{Bechtle:2020uwn}
Philip Bechtle, Sven Heinemeyer, Tobias Klingl, Tim Stefaniak, Georg Weiglein,
  and Jonas Wittbrodt.
\newblock {HiggsSignals-2: Probing new physics with precision Higgs
  measurements in the LHC 13 TeV era}.
\newblock {\em Eur. Phys. J. C}, 81(2):145, 2021, 2012.09197.

\bibitem{ATLAS:2018rnh}
Morad Aaboud et~al.
\newblock {Search for pair production of Higgs bosons in the $b\bar{b}b\bar{b}$
  final state using proton-proton collisions at $\sqrt{s} = 13$ TeV with the
  ATLAS detector}.
\newblock {\em JHEP}, 01:030, 2019, 1804.06174.

\bibitem{Papaefstathiou:2020lyp}
Andreas Papaefstathiou, Tania Robens, and Gilberto Tetlalmatzi-Xolocotzi.
\newblock {Triple Higgs Boson Production at the Large Hadron Collider with Two
  Real Singlet Scalars}.
\newblock {\em JHEP}, 05:193, 2021, 2101.00037.

\bibitem{Barducci:2019xkq}
D.~Barducci, K.~Mimasu, J.~M. No, C.~Vernieri, and J.~Zurita.
\newblock {Enlarging the scope of resonant di-Higgs searches: Hunting for
  Higgs-to-Higgs cascades in $4b$ final states at the LHC and future
  colliders}.
\newblock {\em JHEP}, 02:002, 2020, 1910.08574.

\bibitem{CMS:2018qmt}
Albert~M. Sirunyan et~al.
\newblock {Search for resonant pair production of Higgs bosons decaying to
  bottom quark-antiquark pairs in proton-proton collisions at 13 TeV}.
\newblock {\em JHEP}, 08:152, 2018, 1806.03548.

\bibitem{Robens:2022mvi}
Tania Robens.
\newblock {Models with (broken) $Z_2$ symmetries}.
\newblock {\em PoS}, DISCRETE2020-2021:063, 2022, 2202.09636.

\bibitem{CMS:2022suh}
{Search for a massive scalar resonance decaying to a light scalar and a Higgs
  boson in the four b quarks final state with boosted topology}.
\newblock 4 2022, 2204.12413.

\bibitem{reptr}
Tania Robens.
\newblock {$b\bar{b}b\bar{b}$ final states in the TRSM for asymmetric
  production and decay}.
\newblock https://twiki.cern.ch/twiki/pub/LHCPhysics/LHCHWG3EX/rep.pdf.

\bibitem{Ellwanger:2022jtd}
Ulrich Ellwanger and Cyril Hugonie.
\newblock {Benchmark planes for Higgs-to-Higgs decays in the NMSSM}.
\newblock {\em Eur. Phys. J. C}, 82(5):406, 2022, 2203.05049.

\bibitem{CMS:2021erw}
{Search for a massive scalar resonance decaying to a light scalar and a Higgs
  boson in the four b quark final state with boosted topology}.
\newblock 2021.

\bibitem{CMS:2021fjj}
{Search for new particles in an extended Higgs sector in the four b quark final
  state at $\sqrt{s}=13~\mathrm{TeV}$}.
\newblock 2021.

\bibitem{CMS:2022qww}
Armen Tumasyan et~al.
\newblock {Search for new particles in an extended Higgs sector with four b
  quarks in the final state at $\sqrt{s}$ = 13 TeV}.
\newblock 3 2022, 2203.00480.

\bibitem{ATLAS:2021ypo}
{Search for Higgs boson decays into two spin-0 particles in the $bb\mu\mu$
  final state with the ATLAS detector in $pp$ collisions at $\sqrt{s}=13$ TeV}.
\newblock 3 2021.

\bibitem{ATLAS:2021hbr}
Georges Aad et~al.
\newblock {Search for Higgs boson decays into a pair of pseudoscalar particles
  in the $bb\mu\mu$ final state with the ATLAS detector in $pp$ collisions at
  $\sqrt s$=13\,\,TeV}.
\newblock {\em Phys. Rev. D}, 105(1):012006, 2022, 2110.00313.

\bibitem{EuropeanStrategyforParticlePhysicsPreparatoryGroup:2019qin}
Richard~Keith Ellis et~al.
\newblock {Physics Briefing Book}: {Input for the European Strategy for
  Particle Physics Update 2020}.
\newblock 10 2019, 1910.11775.

\bibitem{European:2720129}
European~Strategy Group.
\newblock {2020 Update of the European Strategy for Particle Physics}.
\newblock Technical report, Geneva, 2020.

\bibitem{Robens:2022zgk}
Tania Robens.
\newblock {A short overview on low mass scalars at future lepton colliders}.
\newblock {\em Universe}, 8:286, 2022, 2205.09687.

\bibitem{Alwall:2011uj}
Johan Alwall, Michel Herquet, Fabio Maltoni, Olivier Mattelaer, and Tim
  Stelzer.
\newblock {MadGraph 5 : Going Beyond}.
\newblock {\em JHEP}, 06:128, 2011, 1106.0522.

\bibitem{ATLAS-CONF-2020-052}
{Combination of searches for invisible Higgs boson decays with the ATLAS
  experiment}.
\newblock Technical report, CERN, Geneva, Oct 2020.
\newblock All figures including auxiliary figures are available at
  https://atlas.web.cern.ch/Atlas/GROUPS/PHYSICS/CONFNOTES/ATLAS-CONF-2020-052.

\bibitem{Drechsel:2018mgd}
P.~Drechsel, G.~Moortgat-Pick, and G.~Weiglein.
\newblock {Prospects for direct searches for light Higgs bosons at the ILC with
  250 GeV}.
\newblock {\em Eur. Phys. J. C}, 80(10):922, 2020, 1801.09662.

\bibitem{Wang:2020lkq}
Yan Wang, Mikael Berggren, and Jenny List.
\newblock {ILD Benchmark: Search for Extra Scalars Produced in Association with
  a $Z$ boson at $\sqrt{s}=500$ GeV}.
\newblock 5 2020, 2005.06265.

\end{thebibliography}

\end{document}